\documentclass[aps,prl,10pt,twocolumn,superscriptaddress,balancelastpage,showpacs,reprint]{revtex4-1}
\usepackage{amssymb,amsmath,amsfonts}
\usepackage{graphicx,color}
\usepackage[squaren]{SIunits}
\usepackage[english]{babel}
\usepackage{hyperref}

\newcommand{\grad}{\nabla}
\renewcommand{\div}{\nabla\cdot}
\renewcommand{\vec}[1]{\mathbf{#1}}

\newcommand\blue[1]{{#1}}
\newcommand\Pe{{\rm Pe}}

\begin{document}
\title{{Generalized Thermodynamics of Phase Equilibria in Scalar Active Matter}}

\author{Alexandre P. Solon}
\thanks{These two authors contributed equally}
\affiliation{Massachusetts Institute of Technology, Department of Physics, Cambridge, Massachusetts 02139, USA}

\author{Joakim Stenhammar}
\thanks{These two authors contributed equally}
\affiliation{Division of Physical Chemistry, Lund University, 221 00 Lund, Sweden}

\author{Michael E. Cates}
\affiliation{DAMTP, Centre for Mathematical Sciences, University of Cambridge, Cambridge CB3 0WA, United Kingdom}

\author{Yariv Kafri}
\affiliation{Department of Physics, Technion, Haifa, 32000, Israel}

\author{Julien Tailleur}
\affiliation{Universit\'e Paris Diderot, Sorbonne Paris Cit\'e, MSC, UMR 7057 CNRS, 75205 Paris, France}

\date{\today}

\begin{abstract}
  Motility-induced phase separation (MIPS) arises generically in
  fluids of self-propelled particles when interactions lead to a
  kinetic slowdown at high densities. {Starting from a continuum
    description of scalar active matter {akin to a generalized
      Cahn-Hilliard equation}, we give a general prescription for the
    mean densities of coexisting phases in flux-free steady states
    that amounts, at a hydrodynamics scale, to extremizing an
    effective free energy}. We illustrate our approach on two
  well-known models: {self-propelled particles interacting either
    through a density-dependent propulsion speed or via direct
    pairwise forces.}  Our theory accounts {quantitatively} for their
  phase diagrams, providing a unified description of MIPS.
\end{abstract}


\maketitle

Active materials, composed of particles individually capable of
dissipatively converting energy into
motion~\cite{Paxton:2004:JACS,Deseigne:2010:PRL,Palacci:2010:PRL,Thutupalli:2011:NJP,Buttinoni:2013:PRL},
display a fascinating range of large-scale
properties~\cite{Ballerini:2008:PNAS,Schaller:2010:PRL,Sumino:2012:Nature,Wensink:2012:PNAS,Marchetti:2013:RMP,Bricard:2013:Nature,Stenhammar:2016:SciAdv}.
Among them, motility-induced phase separation~\cite{Cates:2015:ARCMP}
(MIPS) has recently attracted a lot of
interest~\cite{Tailleur:2008:PRL,Thompson:2011:JSM,Fily:2012:PRL,Redner:2013:PRL,Bialke:2013:EPL,Stenhammar:2013:PRL,Buttinoni:2013:PRL,Wysocki:2014:EPL,Theurkauff:2012:PRL,Soto:2014:PRE,Wittkowski:2014:NC,Brady:2014:PRL,Speck:2014:PRL,Matas:2014:PRE,Zottl:2014:PRL,Suma:2014:EPL,Solon:2015:PRL,Cates:2015:ARCMP,Redner:2016:arxiv}. It
arises because self-propelled particles accumulate in regions where
they move more slowly~\cite{Schnitzer:1993:PRE}. When interactions
between particles lead to their slowing down at high density, a
positive feedback leads to phase separation between a high-density
low-motility phase and a low-density high-motility phase. Remarkably,
this liquid-gas phase separation happens without the need of any
attractive interactions, leading to the emergence of cohesive matter
without cohesive forces. First postulated in idealized toy
models~\cite{Tailleur:2008:PRL,Thompson:2011:JSM,Fily:2012:PRL,Redner:2013:PRL,Bialke:2013:EPL,Stenhammar:2013:PRL,Wysocki:2014:EPL},
MIPS has since been addressed experimentally using self-propelled
colloids~\cite{Theurkauff:2012:PRL,Buttinoni:2013:PRL} and genetically
engineered bacteria~\cite{Liu:2011:SCI}.

The aforementioned instability mechanism leading to MIPS is by now
well understood and has been used to define a spinodal region where
homogeneous phases are linearly
unstable~\cite{Tailleur:2008:PRL,Cates:2015:ARCMP}. Furthermore, this
instability can be understood at the (fluctuating) hydrodynamic
level~\cite{Tailleur:2008:PRL,Cates:2013:EPL,Bialke:2013:EPL,Speck:2014:PRL,Solon:2015:EPJST}
where the dynamics of active particles undergoing a kinetic slowdown
at high density reduce to an equilibrium Model
B~\cite{Hohenberg:1977:RMP}. On the contrary, there is no
comprehensive theory predicting the binodals: the mapping to
equilibrium breaks down at higher order in
gradients~\cite{Wittkowski:2014:NC} and the corresponding equilibrium
predictions for the coexisting binodal densities are
violated~\cite{Wittkowski:2014:NC,Solon:2015:NP}.

MIPS has been observed in two broad classes of systems. In a first
class of
models~\cite{Tailleur:2008:PRL,Thompson:2011:JSM,Cates:2013:EPL,
  Soto:2014:PRE}, MIPS arises from an explicit density-dependence of
the propulsion speed $v(\rho)$. This mimics the way cells adapt their
motion to the local density measured through the concentration of a
chemical signal, and we refer to such particles as `quorum-sensing
active particles' (QSAPs). There, one can define a chemical potential
$\mu$~\cite{Tailleur:2008:PRL} which is equal in coexisting phases,
but the coexisting pressures, whether mechanical~\cite{Solon:2015:NP}
or thermodynamic~\cite{Wittkowski:2014:NC}, are unequal.  In a second
class of
models~\cite{Fily:2012:PRL,Redner:2013:PRL,Bialke:2013:EPL,Stenhammar:2013:PRL},
particles \blue{propelled by a constant force interact via an
isotropic, repulsive pair potential}; the slowdown triggering MIPS is
now due to collisions. Contrary to QSAPs, the mechanical pressure $P$
of such `pairwise-force active particles' (PFAPs), defined as the
force density on a confining wall, is equal in coexisting phases.
However, an effective chemical potential defined from the
thermodynamic equilibrium relation~\cite{Takatori:2015:PRE}
$PV = N\mu-F$ with $\partial F/\partial N = \mu$ takes unequal values
in coexisting phases, causing violation of the equilibrium Maxwell
equal-area construction~\cite{Solon:2015:PRL}. For both models, we
thus lack a constraint to complement the equality of pressure (PFAPs)
or chemical potential (QSAPs) to fix the values of coexisting
densities.  The difference between these two classes of models can be
shown to stem from whether or not an effective momentum conservation
holds in the steady-state~\cite{Fily:2017:JPA}.  When such a
conservation law is present, as in PFAPs, the pressure is given by an
underlying equation of state and is equal in the two
phases~\cite{Brady:2014:PRL,Yang:2014:SM,Solon:2015:PRL}. On the
contrary, in the absence of this conservation law, this is generically
not the case. All in all, a comprehensive theory of the phase
equilibria in MIPS, that would in particular encompass these two
different classes of models, remains elusive.
 
In this Rapid Communication, we propose a unified theory of MIPS based
on phenomenological hydrodynamic equations of motion for the scalar
density field. We show how the binodals are determined at this level
from a common tangent construction on an \textit{effective} free
energy density. Our formalism encompasses equilibrium systems for
which one recovers the standard thermodynamic free energy and, in that
case only, the equality among phases of both pressure and chemical
potential. We then show how this generic formalism can be applied to
precise models of QSAPs and PFAPs, accounting for their phase
diagrams. In particular, we show that different intensive quantities
are equal between coexisting phases in PFAPs and QSAPs.

{\it General framework.} We consider a continuum description of active
particles with isotropic, non-aligning interactions. \blue{In this
  {\it scalar} active matter}, the sole hydrodynamic field is thus the
conserved density $\rho(\vec r,t)$, obeying
$\dot \rho=-\nabla \cdot {\bf J}$.  By symmetry, the current ${\bf J}$
vanishes in homogeneous phases. Its expansion in gradients of the
density involves only odd terms under space reversal. {At third order,
  we use~\footnote{{Note that a generic third order expansion
      ${\bf J}= \alpha \grad\rho-\kappa \grad \Delta \rho+\lambda
      \grad (\grad\rho)^2+[\beta(\grad\rho)^2+\gamma \Delta \rho]
      \grad \rho$
      is formally equivalent to~\eqref{eq:dynamics-general}, at this
      order, using for instance
      $M=1+(\frac\beta\alpha-\frac{\lambda'}{\alpha})(\grad\rho)^2+(\frac\gamma\alpha+\frac{\kappa'}\alpha)\Delta\rho$
      and $g_0$ such that $g_0'(\rho)=\alpha(\rho)$. Here, however, we
      restrict ourselves to Eq.~\eqref{eq:dynamics-general} with a
      positive definite $M$.}}:}
\begin{equation}
  \label{eq:dynamics-general}
\dot\rho= \div ( M \grad g);\;
  g=g_0(\rho) + \lambda(\rho) |\grad \rho|^2-\kappa(\rho)\Delta\rho.
\end{equation}
{The noiseless hydrodynamic equation~\eqref{eq:dynamics-general}
  describes the evolution of the average coarse-grained density field
  on scales much larger than the correlation length and time. It can
  thus be used to characterize fully phase-separated profiles, away
  from the critical point where noise is
  irrelevant~\cite{Bray:2002:AdvPhys}, to predict binodal
  densities. Eq.~\eqref{eq:dynamics-general} plays the same role as
  the Cahn-Hilliard equation does for equilibrium phase-separating
  systems~\cite{Bray:2002:AdvPhys} but in general does not admit an
  equilibrium free energy structure.} In what follows, we first start with
 Eq.~\eqref{eq:dynamics-general} and show how to
compute analytically its phase diagram. We then consider a microscopic
model of QSAPs for which we obtain the coefficients of
Eq.~(\ref{eq:dynamics-general}) in terms of microscopic parameters by
coarse-graining. Finally, we show that our formalism can also be
applied to PFAPs, even though closed expressions of the coefficients
appearing in~\eqref{eq:dynamics-general} are not known explicitly in
this case.

{Equation~\eqref{eq:dynamics-general} predicts a linear instability of
  a homogeneous profile of density $\rho_0$ whenever
  $g_0'(\rho_0)<0$~\footnote{Note that we define $g$ so that $M$ is
    positive}: this is the standard linear instability leading to
  MIPS~\cite{Tailleur:2008:PRL} and defines the spinodal region. We
  now proceed to establish the corresponding binodals. As in
  equilibrium, we consider a fully phase-separated system. A
  macroscopic droplet of the minority phase has an infinite curvature
  radius, and hence effectively flat interfaces, so that curvature
  effects are negligible. As in equilibrium, the problem, though
  $n$-dimensional, reduces to studying the one-dimensional profile
  perpendicular to the interface~\cite{Bray:2002:AdvPhys}. We thus
  consider a flat interface, parallel to $\mathbf{\hat{y}}$, between
  coexisting gas and liquid phases at densities $\rho_g$ and
  $\rho_\ell$.} In a steady state with vanishing current,
$M \grad g=0$, so that $g$ is constant throughout the system{:}
$g[\rho(r,t)]=\bar g$. {This yields} a first equation relating
$\rho_g$ and $\rho_\ell$:
\begin{equation}\label{eq:cond1}
g_0(\rho_g)=g_0(\rho_\ell)=\bar g.
\end{equation}

A second relation can now be obtained by considering a function
$R(\rho)$ and integrating $g(\rho)\partial_x R$ {across} the
interface. Replacing $g(\rho)$ by its value $\bar g$ or its explicit
expression in Eq.~\eqref{eq:dynamics-general}, one finds two
equivalent expressions for
$\int_{x_g}^{x_\ell} g(\rho)\partial_x R \, dx$:
\begin{equation}
  \label{eq:int-mu-R}
  {(R_\ell-R_g)  \bar g}=\phi(R_\ell)-\phi(R_g)+\int_{x_g}^{x_\ell} \!\!\!\!\! [\lambda(\partial_x \rho)^2-\kappa\partial_x^2\rho] \partial_x R\,dx 
\end{equation}
where $x_g$ and $x_\ell$ lie within the bulk gas and liquid phases,
$R_{\ell/g}\equiv R(\rho_{\ell/g})$, and $\phi$ is defined by
$d\phi/dR=g_0(\rho)$. {To simplify Eq.~\eqref{eq:int-mu-R}, we choose
$R(\rho)$ such that
\begin{equation}\label{eqR}
  \kappa R'' = -(2\lambda + \kappa') R',
\end{equation}
where $(')$ denotes $d/d\rho$. Then, one has that}
\begin{equation}
  [\lambda(\partial_x \rho)^2-\kappa\partial_x^2\rho]\partial_xR=-\partial_x\left[\frac{\kappa R'}{2}(\partial_x\rho)^2\right]
\end{equation}
{\blue the integral of which} vanishes between any two bulk planes where
$\partial_x\rho =0$.  Eq.~\eqref{eq:int-mu-R} then yields a second
constraint:
\begin{equation}\label{eq:cond2}
  h_0(R_\ell)=h_0(R_g);\qquad h_0(R) \equiv R \phi'(R)-\phi(R) 
\end{equation}
Because $R$ is nonlinear in $\rho$, the lever rule, $\rho_\ell
V_\ell+\rho_g V_g = \rho_0 V$ is nonlinear in $R$, but still
determines the phase volumes $V_{\ell,g}$. Also the densities
$\rho_{\ell,g}$ do not vary as one moves along the `tie-line' by
changing the global mean density $\rho_0$. This is not true generally
in non-equilibrium phase separation~\cite{Fielding:2003:EPJE}.

{Eqs.}~(\ref{eq:cond1},\ref{eq:cond2}) show the coexisting densities
to satisfy a common tangent construction on an effective (bulk) free
energy $\phi(R) = \int g_0(\rho)dR$. {The mathematical similarity with
an equilibrium common tangent construction can be traced to the fact
that Eq.~\eqref{eq:dynamics-general} can be written  as
\begin{equation}\label{eq:generalizedF}
  \dot \rho= \grad \cdot [M[\rho] \grad g];\qquad g=\frac{\delta {\mathfrak F}}{\delta R},
\end{equation}
with
${\mathfrak F}=\int d {\vec r} [\phi(R) +\frac{\kappa}{2 R'} (\grad
R)^2]$.
The stationary solutions of Eq.~\eqref{eq:dynamics-general} then
correspond to extrema of the `effective' free energy ${\mathfrak
  F}$.
Note that~\eqref{eq:generalizedF} holds in any dimension. This
highlights that, although the construction of the binodals
~\eqref{eq:cond1}-\eqref{eq:cond2} relies on a single coordinate
normal to the interface, our results for the binodals are valid in any
dimensions. Last, since $R(\rho)$ is a bijection, the spinodal region
is equivalently defined by $\phi''(R)<0$ or $g_0'(\rho)<0$.}


To see how our formalism works, let us first consider an equilibrium
  case, in which $g$ has an even simpler form
\begin{equation}\label{eq:FE}
  g=\frac{\delta {\cal F}}{\delta\rho({\bf r})};\quad {\cal F}[\rho] =\int [f(\rho)+\frac{c(\rho)}{2}(\grad\rho)^2] d{\bf r}.
\end{equation}
Eq.~\eqref{eq:dynamics-general} is then the Cahn-Hilliard equation for
a system with free energy ${\cal F}[\rho]$ and mobility
$M[\rho]$~\cite{Cahn:1958:JCP}. Eq.~\eqref{eq:FE} is consistent
with~\eqref{eq:generalizedF} since it imposes $2\lambda+\kappa'=0$ so
that $R=\rho$ \blue{(up to an additive and a multiplicative constant
  which do not affect the phase equilibria)} and
${\mathfrak F}={\cal F}$. We recover $\phi(R)=f(\rho)$ as the bulk
free energy density, $g_0(\rho)=f'(\rho)$ as the chemical potential,
and $h_0(\rho)=f'(\rho) \rho-f(\rho)$ as the pressure.

Our common tangent construction on $\phi(R)$, which amounts to extremizing
${\mathfrak F}$, thus reverts to the usual one in equilibrium,
but extends beyond this. We now show how our formalism can be used to derive the phase
diagrams of QSAPs and PFAPs.

{\it QSAPs.} We consider particles $i=1....N$, moving at speeds
$v_i$ along body-fixed directions $\vec u_i$, which undergo both continuous rotational diffusion
with {diffusivity} $D_r$ and complete randomization with {tumbling
  rate} $\alpha$. Each particle adapts its speed $v(\tilde \rho_i)$ to \blue{the}
local density 
\begin{equation}
  \tilde \rho_i(\vec r)=\int d\vec r' K(\vec r-\vec r')\hat\rho(\vec r') d\vec r'
\end{equation}
with $K(\vec r)$ an
isotropic coarse-graining kernel, and $\hat\rho(\vec r)=\sum_i \delta(\vec
r-\vec r_i)$ the microscopic particle density.

Deriving hydrodynamic equations from microscopics is generally difficult,
even in equilibrium~\cite{Kipnis:2013:book}. For QSAPs we can
follow the path of~\cite{Tailleur:2008:PRL,Cates:2013:EPL,Solon:2015:EPJST},  taking a
mean-field approximation of their \textit{fluctuating}
hydrodynamics. 
We first assume a smooth density field  so that the velocity can be expanded
  as~\cite{supp}
\begin{equation}
  \label{eq:expansion-v}
  v(\tilde\rho_i)\approx v(\rho)+\ell^2v'(\rho)\Delta\rho+{\cal O}(\grad^3)
\end{equation}
where $\rho$ is evaluated at $\vec r_i$ and
$\ell^2=\frac{1}{2}\int r^2 K(\vec r)d\vec r$.
Following~\cite{Cates:2013:EPL,Solon:2015:EPJST}, the fluctuating
hydrodynamics of QSAPs is then given by $\dot\rho=\grad
\cdot (M \grad g + \sqrt{2 M \rho} {\bf\Lambda})$~\cite{supp}, with ${\bf\Lambda}$ a unit white noise vector and
\begin{equation}\begin{aligned}\label{eq:paramMF}
  g_0(\rho)&=\log(\rho v);\quad
  M=\rho\frac{\tau v(\tilde\rho)^2}{d};\\
  \kappa(\rho)&=-\ell^2 \frac{v'}{v};\quad \lambda(\rho)=0\;,
\end{aligned}
\end{equation}
where $d$ is the number of spatial dimensions. Here,
\blue{$\tau\equiv[(d-1)D_r+\alpha]^{-1}$} is the orientational
persistence time. The mean-field hydrodynamic equation of QSAPs is
then Eq.~\eqref{eq:dynamics-general} {with the coefficients in}
Eq.~\eqref{eq:paramMF}. This hydrodynamic description is expected to
hold whenever the correlation length is sufficiently small for the
mean-field approximation to be valid and the interfaces are
sufficiently smooth so that the gradient expansion is justified.

\begin{figure}
  \centering
  \includegraphics[width=0.61\columnwidth]{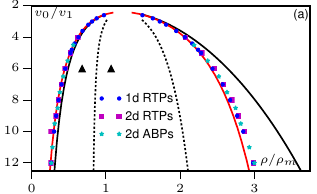}
  \raisebox{1.18cm}{\includegraphics[width=0.375\columnwidth]{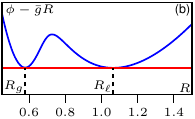}}
  \caption{ {\bf a}: Phase diagrams of QSAPs. The solid lines
    correspond to common tangent constructions on $\phi(R)$ (red) or
    $f(\rho)$ (black). Dashed lines correspond to the spinodals
    \blue{$\phi''(R)=0$}. Data points are from simulations of run-and-tumble
    particles (RTPs, $\alpha=1,\,D_r=0$) or active Brownian particles
    (ABPs, $\alpha=0,\,D_r=1$) in 1D on lattice or 2D in continuous
    space. Black triangles correspond to supplementary movies showing
    nucleation or spinodal decomposition~\cite{supp}. For all plots,
    $v(\rho)=v_0+\frac{v_1-v_0}{2}\left[1+\tanh(\frac{\rho-\rho_m}{L_f})\right]$,
    $K(r)=\exp(-\frac{1}{1-r^2})/Z$ with $Z$ a normalization constant,
    $\rho_m=200$, $v_1=5$, $L_f=100$.  {\bf b}: Common tangent
    construction on $\phi(R)$ for $v_0=20$.}
  \label{fig:vrho}
\end{figure}

{To construct the phase diagram, for a given choice of $v(\rho)$, we
  first solve Eq.~(\ref{eqR}) for $R(\rho)$ and use it to obtain both
  $\phi(R)$ and $h_0(R)$. The binodals then follow via a
  common-tangent construction on $\phi(R)$ or, equivalently, by
  setting equal values of $h_0$ and $g_0$ in {coexisting} phases. Note
  that since $2\lambda+\kappa'\neq 0$ one has $R \neq \rho$.} 


Fig.~\ref{fig:vrho} shows the phase diagrams predicted by our
generalized thermodynamics and by QSAP simulations.  As expected, the
hydrodynamic description works best fairly close to the critical point
(but outside a numerically unresolved Ginzburg interval where
fluctuations cannot be neglected). This is where interfaces are
smoothest and the gradient expansion Eq.~\eqref{eq:expansion-v} most
accurate.  To determine precisely the binodals, we choose a $v(\rho)$
(given in the caption of Fig.~\ref{fig:vrho}) such that MIPS occurs
only at large densities, leading to well-separated coexisting
densities.  Under these conditions, our mean-field approximation works
very well: the agreement between predicted and measured binodals is
excellent. In contrast, a common tangent construction on $f(\rho)$
defined by $f'(\rho)=g_0(\rho)$ as proposed
before~\cite{Tailleur:2008:PRL,Cates:2013:EPL} gives a poorer estimate
since it correctly captures the equality of $g_0$ in both phases but
not that of $h_0$. This reminds us that {\em gradient terms directly
  influence the coexisting densities} through Eq.~\eqref{eqR} -- quite
unlike the equilibrium case.  As an aside, it is remarkable that for
QSAPs we can quantitatively predict the phase diagram of a microscopic
model without any fitting parameter, something rare even for
equilibrium models.

Beyond the quantitative prediction of the phase diagram, our approach
{provides} insight into the universality of the MIPS seen for
QSAPs. For instance, the phase diagram does not depend on the kernel
$K$, which enters Eq.~\eqref{eq:paramMF} only through the constant
$\ell^2$ which then cancels from Eq.~(\ref{eqR}) defining the nonlinear
transform $R(\rho)$.  Likewise, Fig.~\ref{fig:vrho} includes lattice
simulations of QSAPs in $1d$ where full phase separation is replaced
by alternating domains (whose densities obey the predicted binodal
values), and confirms the equivalence of continuous (ABP) and discrete
(RTP) angular relaxation dynamics for QSAPs
\cite{Cates:2013:EPL,Solon:2015:EPJST}.

{\it PFAPs.} We now consider self-propelled particles, of diameter $\sigma$, 
in $2d$, interacting via a short-range
repulsive pair potential $V$ (see~\cite{supp} for details):
\begin{equation}
  \dot {\bf r}_i=-\sum_j \grad_i V(|{\bf r}_i-{\bf r}_j|)+ \sqrt{2D_t} {\boldsymbol \xi}_i+v_0 {\bf u}_i;\;\dot \theta_i=\sqrt{2D_r}\eta_i.\nonumber
\end{equation}
Here a microscopic mobility multiplying the first term was set to
unity; ${\bf u}_i=(\cos\theta_i,\sin\theta_i)$, and
$\eta_i,{\boldsymbol \xi}_i$ are unit Gaussian white noises. For
simplicity, we only include continuous rotational diffusion, but we
expect our results to stand for tumbles as well since this difference
as been shown to have a negligible effect on the phase
equilibria~\cite{Solon:2015:EPJST}. MIPS occurs in this system if the
{P\'eclet} number ${\rm Pe}=3v_0/(\sigma D_r)$ exceeds a threshold value
${\rm Pe}_c\sim
60$~\cite{Fily:2012:PRL,Redner:2013:PRL,Bialke:2013:EPL,Stenhammar:2013:PRL}.

We follow~\cite{farrell:2012:PRL,Solon:2015:PRL} to derive a
fluctuating hydrodynamics for the stochastic density
$\hat\rho({\bf r})=\sum_{i=1}^N \delta({\bf r}-{\bf r}_i)$ \blue{the
  deterministic limit of which} gives a coarse-grained equation for
the mean density field. On time scales larger than $D_r^{-1}$, in our
phase-separated set-up with a {flat} interface parallel to
{$\mathbf{\hat{y}}$}, the dynamics is given by
$  \dot\rho=\partial_{x}^2 g$~\cite{Solon:2015:PRL},
with
\begin{eqnarray}\label{pressure-field}
  g([\rho],x) &=& {D_t}\rho+\frac{v_0^2}{2D_r}(\rho+m_2)+ \hat I_2-\frac{v_0 D_t}{D_r} \partial_x m_1+P_{D};\nonumber\\
  P_D &=& \int_{-\infty}^x dx \int \partial_xV(\vec r'-\vec r)\langle \hat \rho(\vec r')\hat \rho(\vec r)\rangle d^2{\vec r}'; \\
  \hat I_2 &=&-\frac{v_0}{D_r}\int \partial_x V(\vec r'-\vec r)\langle \hat \rho(\vec r')\hat m_1(\vec r)\rangle d^2{\vec r}'. \nonumber
\end{eqnarray}
Here, $\hat{m}_n= \sum_{i=1}^N \delta({\bf r}-{\bf r}_i) \cos
(n\theta_i)$ and $m_n=\langle \hat m_n \rangle$, where
$\langle\dots\rangle$ represent averages over noise realizations.
The lack of steady-state current shows $g$ to be uniform in the
  phase-separated system, equal to some constant $\bar g$.
  
For homogeneous systems the expression for $g$ in
Eq.~\eqref{pressure-field} reduces exactly to the equation of state
(EOS) found previously for the mechanical pressure $P$ of
PFAPs~\cite{Solon:2015:PRL}. Thus $g$ is equal between phases, as it
was for QSAPs, but now it represents pressure, not chemical
potential. Moreover, Eq.~\eqref{pressure-field} generalizes the
pressure EOS of~\cite{Solon:2015:PRL} to inhomogeneous situations. It
can formally be written $g = g_0(\rho(x)) + g_{\rm int}([\rho],x)$
where $g_0(\rho)$ is the pressure in a notionally homogeneous system
with average density $\rho$, and the `interfacial' term $g_{\rm int}$ represents
all nonlocal corrections to this. The form of $g$ used in
Eq.~\eqref{eq:dynamics-general} can then be viewed as a
gradient expansion of Eq.~\eqref{pressure-field} for PFAPs.

One way forward would be to make that expansion (or perhaps avoid it
by using a closed-form ansatz for $g_{\rm int}$), and then find
$R(\rho)$ and $\phi(R)$ analytically as was done for QSAPs above. Here
however we proceed differently, approximating instead the local part,
$g_0(\rho)$, of $g$ in Eq.~\eqref{pressure-field} by a well
benchmarked, {semi-empirical} EOS, with parameters constrained by
simulations of uniform phases at $\Pe=40<\Pe_c$~\cite{supp}. We thus
retain the {\em exact structure} of the nonlocal terms,
$g_{\rm int}(x) \equiv g([\rho],x)-g_0(\rho(x))$ in
Eq.~\eqref{pressure-field}, but find them numerically.  Although less
predictive than knowing such terms algebraically, our method clearly
illustrates how they select the binodals. Furthermore, $g_{\rm int}$
includes all orders in gradient and hence does not rely on a gradient
expansion, contrary to Eq.~\eqref{eq:dynamics-general}.

We then proceed as in Eq.~\eqref{eq:int-mu-R} \blue{but, instead of $R$,}
now using the volume per particle $\nu \equiv \rho^{-1}$. The integral
$\int_{x_g}^{x_\ell} (g-g_0) \,\partial_x \nu \, dx$ then admits two
equivalent expressions
\begin{eqnarray}
  \label{eq:ABPs-unequal-area}
  \int_{\nu_\ell}^{\nu_g}(g_0(\nu)-\bar g)d\nu\label{GMC1}=
  \int_{x_g}^{x_\ell}g_{\rm int}\,\partial_x \nu\, dx
   \label{GMC2}.
 \end{eqnarray}
 {Here $g_0(\nu)$ is the pressure-volume EOS, so that the non-zero
   value of the right hand integral {\em directly} quantifies
   violation of the Maxwell construction.  A fully predictive theory
   would evaluate the right hand side integral and then solve
   $ g_0(\nu_\ell)= g_0(\nu_g)=\bar g$ together with
   Eq.~\eqref{eq:ABPs-unequal-area} to obtain the values of the
   binodals $\nu_\ell$ and $\nu_g$.  In practice, {we measure $g(x)$
     numerically} via Eq.~\eqref{pressure-field} from which we
   subtract $g_0(\rho(x))$ and integrate over space to obtain the
   numerical value of the right hand side of
   Eq.~\eqref{eq:ABPs-unequal-area}. Crucially, $\bar g$, $\nu_g$ and
   $\nu_\ell$ are not inputs here, but are found by solving
   Eq.~\eqref{eq:ABPs-unequal-area}. Concretely this is done via a
   modified Maxwell construction: The binodals correspond to the
   intersect between the function $g_0(\nu)$ and a horizontal line of
   (unknown) ordinate $\bar g$ since
   $g_0(\nu_\ell)=g_0(\nu_g)=\bar g$. We then adjust the value of
   $\bar g$ to solve Eq.~\eqref{eq:ABPs-unequal-area}. This
   construction is illustrated in Fig.~\ref{fig:ABPs}, and is
   accurately obeyed by simulations, unlike the equilibrium Maxwell
   construction, which (notwithstanding \cite{Takatori:2015:PRE})
   clearly fails to account for the phase equilibria of PFAPs where
   interfacial terms again directly enter.

\begin{figure}
  \centering
  \hspace{-.3cm}\includegraphics[width=0.95\columnwidth]{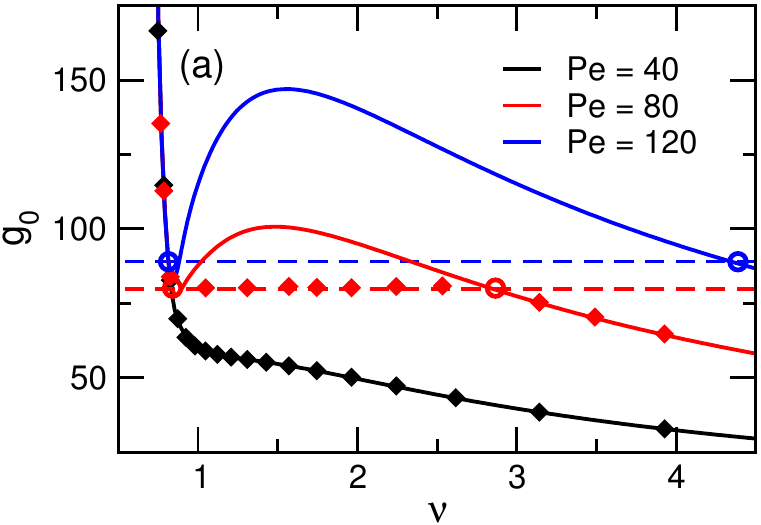}
    \includegraphics[width=0.95\columnwidth]{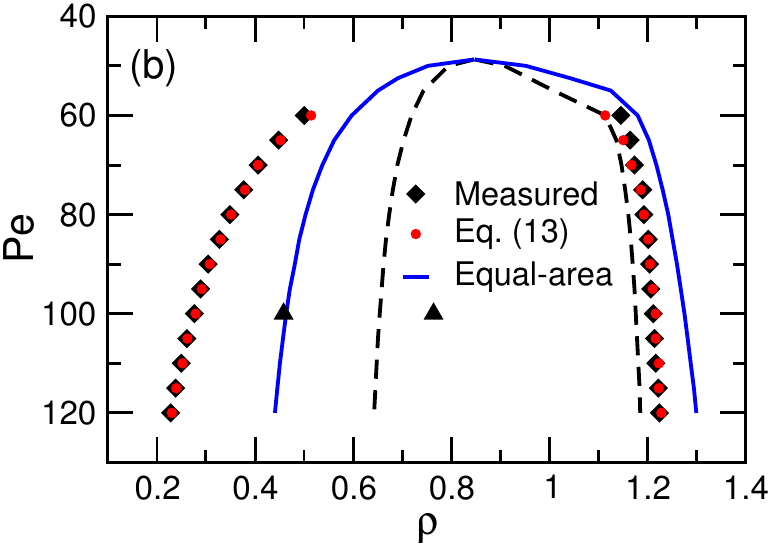}
  \caption{{\bf a}: {Mechanical pressure of PFAPs. Semi-empirical EOS
      $g_0(\nu)$ (line) vs numerical measurements (symbols) for
      various $\Pe$. Open symbols correspond to binodals, horizontal
      lines to the pressure $\bar g$ predicted by Eq.~\eqref{GMC2}. }
    {\bf b}: {Corresponding} phase diagrams obtained via the modified
    (red; see text) and the equal-area (blue) Maxwell constructions,
    compared with numerically measured binodals (black).  {Dashed
      lines correspond to the spinodals $g_0'(\rho)=0$. Black
      triangles correspond to supplementary movies showing nucleation
      or spinodal decomposition~\cite{supp}.}}
  \label{fig:ABPs}
\end{figure}

In this article, we have shown how to build a generalized {theory} of
phase-separating scalar active matter starting from a generalized
Cahn-Hilliard description derived on symmetry grounds. Our work
accounts for the phase equilibria of two {important} classes of
self-propelled particles, PFAPs and QSAPs, which each undergo MIPS. In
contrast to equilibrium systems, interfacial contributions to pressure
and/or chemical potential generically affect the binodal densities at
coexistence \cite{Solon:2015:PRL,Wittkowski:2014:NC}.  We have given
in Eqs.~(\ref{eq:cond1},\ref{eq:cond2}) an {\em explicit construction}
for the binodals at leading non-trivial order in a gradient expansion. This is
quantitatively accurate for MIPS in QSAPs at high density. In
Eq.~\eqref{eq:ABPs-unequal-area} we have given a more general
construction that holds beyond the gradient expansion; we tested it
using numerical data on PFAPs. In practice, our results are obtained
by deriving the coexisting densities of fully phase-separated profiles
in the steady-state. Extending our formalism to account for the
dynamical convergence to this state is an exciting challenge left for
future works. Similarly, the fate of our generalized thermodynamic
formalism when more than one conserved field is present is an open
question.

Interestingly, QSAPs and PFAPs share the same mathematical structure
but their coexisting densities are selected by equating intensive
observables which have different physical interpretation. In
particular, the mechanical pressure is identical in coexisting phases
for PFAPs, but not for QSAPs, due to the lack of an effective momentum
conservation in the latter case~\cite{Fily:2017:JPA}. This fundamental
difference is well captured by our formalism which indeed leads to
different observables $g$ for the two models.

Beyond {understanding} the phase equilibria of active matter, we hope that
our approach will pave the way towards a more general definition of
intensive {thermodynamic}
parameters~\cite{Bertin:2006:PRL,Bertin:2007:PRE,Dickman:2016:NJP} for
active systems. {Building a thermodynamic theory of active matter would
  further improve our understanding and control of these intriguing
  systems and has become a central question in the
  field~\cite{Tailleur:2008:PRL,Palacci:2010:PRL,Brady:2014:PRL,Yang:2014:SM,Speck:2014:PRL,Wittkowski:2014:NC,Solon:2015:EPJST,Solon:2015:PRL,Solon:2015:NP,Ginot:2015:PRX,Takatori:2015:PRE,Bialke:2015:PRL,Farage:2015:PRE,Marconi:2015:SM,Fodor:2016:PRL,Dijkstra:2016:arxiv}.}

\nocite{Plimpton:1995:JCompPhys,Allen:1989,Dean:1996:JPA}

{\em Acknowledgements:} We thank M. Kardar, H. Touchette for
discussions.  APS acknowledges funding through a PLS fellowship from
the Gordon and Betty Moore foundation. JS is funded by a Project Grant
from the Swedish Research Council (2015-05449). MEC is funded by the
Royal Society. This work was funded in part by EPSRC Grant EP/J007404.
YK is supported by an I-CORE Program of the Planning and Budgeting
Committee of the Israel Science Foundation and an Israel Science
Foundation grant. JT acknowledges the support of ANR grant
Bactterns. JT and YK acknowledge a joint CNRS-MOST travel grant. The
PFAP simulations were performed on resources provided by the Swedish
National Infrastructure for Computing (SNIC) at LUNARC.

\bibliographystyle{apsrev4-1}
\bibliography{biblio-MIPS}
\end{document}